\documentstyle[12pt,epsf]{article}
\begin{document}
\baselineskip 0.8cm

\newcommand{\gsim}{ \mathop{}_{\textstyle \sim}^{\textstyle >} }
\newcommand{\lsim}{ \mathop{}_{\textstyle \sim}^{\textstyle <} }
\newcommand{\vev}[1]{ \left\langle {#1} \right\rangle }
\newcommand{\EV}{ {\rm eV} }
\newcommand{\KEV}{ {\rm keV} }
\newcommand{\MEV}{ {\rm MeV} }
\newcommand{\GEV}{ {\rm GeV} }
\newcommand{\TEV}{ {\rm TeV} }
\def\tr{\mathop{\rm tr}\nolimits}
\def\Tr{\mathop{\rm Tr}\nolimits}
\def\Re{\mathop{\rm Re}\nolimits}
\def\Im{\mathop{\rm Im}\nolimits}
\setcounter{footnote}{1}

\begin{titlepage}

\begin{flushright}
UT-856\\
\end{flushright}

\vskip 2cm
\begin{center}
{\large \bf  Cosmological Constants as Messenger between Branes}
\vskip 1.2cm
Izawa K.-I.$^{1,2}$, Yasunori Nomura$^1$, and T.~Yanagida$^{1,2}$

\vskip 0.4cm

$^{1}$ {\it Department of Physics, University of Tokyo, \\
         Tokyo 113-0033, Japan}\\
$^{2}$ {\it Research Center for the Early Universe, University of Tokyo,\\
         Tokyo 113-0033, Japan}
\vskip 1.5cm

\abstract{
 We present a supersymmetry-breaking scenario in which both the breaking
 in the hidden sector with no-scale type supergravity and that in the
 observable sector with gauge mediation are taken into account. 
 The breaking scales in the hidden and observable sectors are related
 through the vanishing condition of the cosmological constant with a
 brane-world picture in mind.
 Suppressing flavor-changing neutral currents, we can naturally obtain
 the gravitino, Higgs(ino), and soft masses of the electroweak scale.
}

\end{center}
\end{titlepage}

\renewcommand{\thefootnote}{\arabic{footnote}}
\setcounter{footnote}{0}

%
%
%       *** Main Part ***
%
%

In supergravity, supersymmetry(SUSY)-breaking effects in a (so-called)
hidden sector are transmitted into the observable sector through
nonrenormalizable interactions \cite{SUGRA}.
With a generic K\"{a}hler potential, we have arbitrary soft SUSY-breaking 
masses for squarks and sleptons, which generate too large flavor-changing 
neutral currents(FCNC's) at low energies \cite{FCNC}.
A solution to this problem is to consider that the hidden sector
responsible for the SUSY breaking is fully separated from the observable 
sector not only in the superpotential $W$ but also in the K\"{a}hler
potential $K$.
However, physical contents of the separation of two sectors
depend strongly on the frames we take in supergravity.

The most popular separation is given in the Einstein frame \cite{SUGRA}, 
which generates a common SUSY-breaking mass for all squarks and
sleptons.
This degeneracy in the soft masses suppresses sufficiently the unwanted
FCNC's \cite{FCNC}.
Although the separation in the Einstein frame is well consistent with
experimental constraints, the origin of the separation is not clear enough.

An alternative has been proposed in Ref.~\cite{IKYY}
which assumes the separation in the ``conformal'' frame in supergravity.
The K\"{a}hler potential ${\cal K}$ and superpotential $W$ are
postulated to have the following forms \cite{IKYY}:
\begin{eqnarray}
  {\cal L} = \int d^2\Theta\, 2{\cal E}
    \left[ -\frac{1}{8}(\bar{\cal D}\bar{\cal D}-8R)\,
           {\cal K}(Q, Q^{\dagger}, Z, Z^{\dagger})
           + W(Q, Z) \right] + {\rm h.c.};
\end{eqnarray}
\begin{eqnarray}
  {\cal K}(Q, Q^{\dagger}, Z, Z^{\dagger})
     &=& -3 + f_O(Q, Q^{\dagger}) + f_H(Z, Z^{\dagger}), \\
  W(Q, Z) &=& W_O(Q) + W_H(Z).
\end{eqnarray}
Here, $Q$ and $Z$ denote fields in the observable and hidden sectors,
respectively.
In the Einstein frame, we see that the K\"{a}hler potential $K$ has the
form of no-scale type\footnote{
The no-scale supergravity \cite{no-scale} adopts a specific form 
$f_H = Z + Z^{\dagger}$.}
as
\begin{eqnarray}
  K(Q, Q^{\dagger}, Z, Z^{\dagger}) = -3\log
    \left(1 - \frac{1}{3}f_O(Q, Q^{\dagger}) 
    - \frac{1}{3}f_H(Z, Z^{\dagger})\right).
\end{eqnarray}

It is interesting that the above separation of the hidden and observable
sectors is stable against radiative corrections due to matter-field
loops, and we may expect some underlying physics that naturally explains 
the separation.
Recently, Randall and Sundrum \cite{RS1} have suggested a beautiful
geometric explanation on the hidden and observable separation in the
``conformal'' frame.
They have claimed that the hidden and observable sectors live on
different three-dimensional ``branes'' separated by a gravitational bulk
\cite{Hor}
in higher-dimensional spacetime.
Although the details of the seraration are not fully clarified,
the picture of geometric separation may deserve further investigation.

It is a crucial observation in Ref.~\cite{IKYY} that all soft
SUSY-breaking masses and $A$ terms in the observable sector vanish in
the limit of the zero cosmological constant.
All gaugino masses in the observable sector also vanish because of the
decoupling of hidden field $Z$ from the gauge kinetic function
\cite{RS1}.
The $B$ term of Higgs fields is exclusively
the soft SUSY-breaking parameter in 
the observable sector, which arises from the $F$ component of an
auxiliary field $\Phi$ of the gravitational supermultiplet \cite{RS1}.
The soft SUSY-breaking mass $B$ should be chosen as $B \lsim 1~\TEV$ to
cause naturally the electroweak symmetry breaking at $O(100)~\GEV$.
This requires the gravitino mass $m_{3/2} \lsim 1~\TEV$, and we take
here $m_{3/2} \sim 1~\TEV$.
In these circumstances, the anomaly mediation \cite{RS1,GLMR} generates
too small SUSY-breaking masses in the observable sector and hence the
gauginos, squarks, and sleptons remain almost massless, rendering the
proposals in Ref.~\cite{IKYY,RS1} unsatisfactory.

The above argument leads us to consider another source of SUSY breaking
in the observable sector and postulate gauge mediation
\cite{gauge_med} yielding sufficiently large SUSY-breaking masses for
the SUSY standard-model particles.\footnote{
Mixture of gravity and gauge mediations is considered in
Ref.~\cite{Pop}.}
However, this scenario has a manifest drawback: there is no reason
why the scale of SUSY-breaking masses (of order $100~\GEV - 1~\TEV$) in the
observable sector coincides with the gravitino mass $m_{3/2}$ arising
from the hidden sector, since the two branes corresponding to the two
sectors are separated by a bulk and the dynamics on each brane
are most likely independent.

In this paper, we consider a possible scenario where the dynamical
scale on one brane is strongly related to the scale on the other brane
in order to cancel vacuum energies produced on the two branes.
Namely, nonvanishing ``cosmological constants'' appearing on the two branes
play a role of messenger between the hidden and observable sectors.
This is plausible, for instance, if the vanishing cosmological constant
in four dimensions is achieved for some higher-dimensional reasons
\cite{Rub}
due to the presence of the bulk.

Let us consider a situation that the scale of the SUSY breaking
arises dominantly on the hidden brane:
the $F$ component of a hidden chiral superfield $Z$ is determined as 
$|\vev{F_Z}| \simeq m_{3/2}M_{\rm pl} \sim (10^{10.5}~\GEV)^2$ so that
we get $m_{3/2} \sim 1~\TEV$.
We can fix the condensation of the superpotential
to cancel the positive cosmological constant arising
from the hidden-sector SUSY breaking:
\begin{eqnarray}
  |\vev{F_Z}|^2 - 3\frac{|\vev{W}|^2}{M_{\rm pl}^2} = 0.
\end{eqnarray}
Note that the bulk contribution to the vacuum condensation
is implicit in the superpotential and determined by the
higher-dimensional equations of motion \cite{RS2}.
Having higher-dimensional reasons for the vanishing cosmological
constant in mind, we suspect that the vacuum energy does not cancel out
predominantly within the hidden brane. 
Thus, we assume that the scale of the superpotential condensation is
given mainly by the dynamics on the observable brane.\footnote{
This does not necessarily mean that the bulk cosmological constant is
negligible.}
Then, we can determine the dynamical scale
$\Lambda$ in the observable sector responsible for the
condensation as
$\Lambda \sim (|\vev{F_Z}|M_{\rm pl})^{1/3} 
\sim (m_{3/2}M_{\rm pl}^2)^{1/3} 
\sim 10^{13}~\GEV$.

For concreteness, we consider an observable-sector chiral superfield $S$ 
carrying $R$-charge $2/3$
whose condensation is given by $\langle S \rangle = \Lambda$.
This is realized, for instance, by a dynamical condensation
of the matter $Q{\bar Q}$ considered in
Ref.\cite{Hot,Iza,INTY}
with $S$ given by $Q{\bar Q}/M_{\rm pl}$.
We have an $R$-symmetric superpotential as
\begin{eqnarray}
  W_O \supset f S^3,
\end{eqnarray}
which induces the superpotential condensation of an appropriate size.

The condensation affects various aspects of
other components in the observable sector, since they are directly
connected in the superpotential.
We now discuss an example of possible effects.
Let us adopt a SUSY SU(2) gauge theory with $2N_f$ doublet hyperquarks
$Q_{\alpha}^{\prime\,i}$ $(\alpha=1,2;\, i=1,\cdots,2N_f)$ for a
demonstration of our point.
We assume that the first four hyperquarks $Q_{\alpha}^{\prime\,i}$
$(i=1,\cdots,4)$ carry vanishing $R$ charges, while the remaining hyperquarks
$Q_{\alpha}^{\prime\,i}$ $(i=5,\cdots,2N_f)$ carry $R$-charge $2/3$.
Then, the latter hyperquarks $Q_{\alpha}^{\prime\,i}$ $(i=5,\cdots,2N_f)$
acquire masses of the order of the $R$-breaking scale $\Lambda$ through
the following superpotential:
\begin{eqnarray}
  W_O \supset h_{ij} Q^{\prime\,i} Q^{\prime\,j} S.
\end{eqnarray}
Therefore, we have a SUSY SU(2) gauge theory with four massless
hyperquarks $Q_{\alpha}^{\prime\,i}$ $(i=1,\cdots,4)$ below the
$R$-breaking scale $\Lambda$, and the gauge coupling becomes strong
causing nontrivial dynamics at a lower energy scale 
$\Lambda^{\prime} \sim (\Lambda_{N_f}^{6-N_f} \Lambda^{N_f-2})^{1/4}$.
Here, $\Lambda_{N_f}$ denotes the dynamical scale determined by
the SU(2) gauge coupling well above the scale $\Lambda$.
This implies that $\Lambda'$ is expected to be close to the $R$-breaking
scale $\Lambda$ for appropriate values of $N_f$
\cite{Iza}.
This SU(2) gauge theory can be arranged to break SUSY dynamically.
In fact, an introduction of six singlets $S_{ij}$ $(= -S_{ji})$
$(i,j=1,\cdots,4)$ which couple to $Q^{\prime\,i}Q^{\prime\,j}$ in the
superpotential generates dynamical SUSY breaking \cite{IYIT}, whose
scale is given by the scale $\Lambda'$ of the SU(2) theory.

Postulating suitable messenger fields \cite{Hot,Iza,INTY,NTY},
the above SUSY
breaking effects are transmitted into the SUSY standard-model sector.
Analyses in Ref.~\cite{Hot,Iza,INTY,NTY} suggest
$\Lambda' \sim 10^5~\GEV-10^9~\GEV$,
which seems a quite reasonable value when one compares it with the
$R$-breaking scale $\Lambda \sim 10^{13}~\GEV$.
This is analogous to the relation between the QCD and the electroweak
symmetry-breaking scales.

We should note here that the present scenario solves two serious
problems in the genuine gauge mediation.
First of all, there is no $\mu$ problem.
The $\mu$ term (SUSY-invariant mass for Higgs multiplets $H$ and
$\bar{H}$) arises naturally from the $R$-symmetry breaking 
\cite{IKYY, mu}.
Namely, the Higgs supermultiplets $H$ and $\bar{H}$ couple to the $S$
field, provided they have vanishing $R$ charges, in the superpotential
as
\begin{eqnarray}
  W_O \supset \frac{k}{M_{\rm pl}^2} H \bar{H} S^3,
\end{eqnarray}
which induces
$\mu \simeq k \vev{S}^3 / M_{\rm pl}^2 \simeq 100~\GEV-1~\TEV$.
Second, the gravitino problem is less severe, since 
$m_{3/2} \sim 1~\TEV$.
The inflationary universe with reheating temperature 
$T_R \simeq 10^8~\GEV$ is consistent with the big-bang nucleosynthesis
for $m_{3/2} \simeq 1~\TEV$ \cite{gravitino}.
It has been shown \cite{AHKK}, recently, that with 
$T_R \simeq 10^8~\GEV$ the leptogenesis works very well.

We have assumed, so far, the minimal SUSY standard model as for the
usual quark, lepton, and Higgs fields.
However, if we adopt the next-to-minimal SUSY standard model
\cite{NMSSM}, the soft SUSY-breaking $B$ term does not appear directly
and hence we may raise the gravitino mass up to
$m_{3/2} \sim 100~\TEV$.\footnote{
Several models are proposed \cite{Pom} which incorporate this heavy
gravitino.}
In this case, the anomaly-mediation effects \cite{RS1,GLMR} may
dominate the gaugino masses giving rise to interesting experimental
signals \cite{sign}, if the gaugino masses induced by the gauge
mediation are suppressed as in Ref.~\cite{INTY}.
The problem of tachyonic sleptons \cite{RS1} in the anomaly mediation
may be naturally solved by the gauge mediation discussed in this paper.

\vspace{7mm}

{\bf Acknowledgments}

Y.N. thanks the Japan Society for the Promotion of Science for financial 
support.
This work was partially supported by ``Priority Area: Supersymmetry and
Unified Theory of Elementary Particles (\# 707)'' (T.Y.).

\newpage
%
%%%%%%%%%%%%%%%%%%%%%%%%%%%%%%%%%%%%%%%%%%%%%%%%%%%%%%%%%%%%%%%
%
% NEW COMMANDS FOR THE BIBLIOGRAPHY
%
%%%%%%%%%%%%%%%%%%%%%%%%%%%%%%%%%%%%%%%%%%%%%%%%%%%%%%%%%%%%%%%
\newcommand{\Journal}[4]{{\sl #1} {\bf #2} {(#3)} {#4}}
\newcommand{\PL}{\sl Phys. Lett.}
\newcommand{\PR}{\sl Phys. Rev.}
\newcommand{\PRL}{\sl Phys. Rev. Lett.}
\newcommand{\NP}{\sl Nucl. Phys.}
\newcommand{\ZP}{\sl Z. Phys.}
\newcommand{\PTP}{\sl Prog. Theor. Phys.}
\newcommand{\NC}{\sl Nuovo Cimento}
\newcommand{\MPL}{\sl Mod. Phys. Lett.}
\newcommand{\PRep}{\sl Phys. Rep.}
%%%%%%%%%%%%%%%%%%%%%%%%%%%%%%%%%%%%%%%%%%%%%%%%%%%%%%%%%%%%%%%

%

\begin{thebibliography}{99}
%
\bibitem{SUGRA}
    For a review, H.P.~Nilles,
        {\PRep} {\bf 110} (1984) 1.
%
\bibitem{FCNC}
    F.~Gabbiani, E.~Gabrielli, A.~Masiero, and L.~Silvestrini,
        {\NP} {\bf B477} (1996) 321.
%
\bibitem{IKYY}
    K.~Inoue, M.~Kawasaki, M.~Yamaguchi, and T.~Yanagida,
        {\PR} {\bf D45} (1992) 328.
%
\bibitem{no-scale}
    For a review, A.~Lahanas and D.V.~Nanopoulos,
        {\PRep} {\bf 145} (1987) 1.
%
\bibitem{RS1}
    L.~Randall and R.~Sundrum,
        hep-th/9810155.
%
\bibitem{Hor}
    P.~Horava and E.~Witten,
        {\NP} {\bf B460} (1996) 506; 
        {\NP} {\bf B475} (1996) 94;\\
    E.~Witten, {\NP} {\bf B471} (1996) 135.
%
\bibitem{GLMR}
    G.F.~Giudice, M.A.~Luty, H.~Murayama, and R.~Rattazzi,
        {\sl JHEP} {\bf 9812} (1998) 027.
%
\bibitem{gauge_med}
    For reviews, G.F.~Giudice and R.~Rattazzi,
        hep-ph/9801271;\\
    S.L.~Dubovsky, D.S.~Gorbunov, and S.V.~Troitsky,
        hep-ph/9905466.
%
\bibitem{Pop}
    E.~Poppitz and S.P.~Trivedi,
        {\PR} {\bf D55} (1997) 5508. 
%
\bibitem{Rub}
    V.A.~Rubakov and M.E.~Shaposhnikov,
        {\PL} {\bf B125} (1983) 139.
%
\bibitem{RS2}
    L.~Randall and R.~Sundrum,
        hep-ph/9905221;\\
    H.~Verlinde, 
        hep-th/9906182;\\
    A.~Kehagias, 
        hep-th/9906204.
%
\bibitem{Hot}
    T.~Hotta, Izawa~K.-I., and T.~Yanagida,
        {\PR} {\bf D55} (1997) 415.
%
\bibitem{Iza}
    Izawa~K.-I.,
        {\PTP} {\bf 98} (1997) 443.
%
\bibitem{INTY}
    Izawa~K.-I., Y.~Nomura, K.~Tobe, and T.~Yanagida,
        {\PR} {\bf D56} (1997) 2886;\\
    Y.~Nomura and K.~Tobe,
        {\PR} {\bf D58} (1998) 055002.
%
\bibitem{IYIT}
    Izawa~K.-I. and T.~Yanagida,
        {\PTP} {\bf 95} (1996) 829;\\
    K.~Intriligator and S.~Thomas,
        {\NP} {\bf B473} (1996) 121.
%
\bibitem{NTY}
    Y.~Nomura, K.~Tobe, and T.~Yanagida,
        {\PL} {\bf B425} (1998) 107.
%
\bibitem{mu}
    T.~Yanagida,
        {\PL} {\bf B400} (1997) 109.
%
\bibitem{gravitino}
    E.~Holtmann, M.~Kawasaki, K.~Kohri, and T.~Moroi,
        {\PR} {\bf D60} (1999) 023506,
    and references therein.
%
\bibitem{AHKK}
    T.~Asaka, K.~Hamaguchi, M.~Kawasaki, and T.~Yanagida,
        hep-ph/9907559.
%
\bibitem{NMSSM}
    J.~Ellis, J.F.~Gunion, H.E.~Haber, L.~Roszkowski, and F.~Zwirner,
        {\PR} {\bf D39} (1989) 844;\\
    A.~de Gouvea, A.~Friedland, and H.~Murayama,
        {\PR} {\bf D57} (1998) 5676;\\
    and references therein.
%
\bibitem{Pom}
    A.~Pomarol and R.~Rattazzi,
        {\sl JHEP} {\bf 9905} (1999) 013;\\
    Z.~Chacko, M.A.~Luty, I.~Maksymyk, and E.~Ponton,
        hep-ph/9905390;\\
    E.~Katz, Y.~Shadmi, and Y.~Shirman,
        hep-ph/9906296.
%
\bibitem{sign}
    J.L.~Feng, T.~Moroi, L.~Randall, M.~Strassler, and S.~Su,
        hep-ph/9904250;\\
    T.~Gherghetta, G.F.~Giudice, and J.D.~Wells,
        hep-ph/9904378;\\
    J.F.~Gunion and S.~Mrenna,
        hep-ph/9906270;\\
    J.L.~Feng and T.~Moroi,
        hep-ph/9907319.
%%
\end{thebibliography}
\end{document}